\documentclass{acmsiggraph}

\usepackage{balance} 
\usepackage{graphics} 
\usepackage{times}   
\usepackage{url}  
\usepackage[activate]{pdfcprot}
\usepackage[usenames,dvipsnames]{xcolor}
\usepackage{tabularx}
\usepackage{changes}
\usepackage{booktabs}
\usepackage{bm}
\usepackage{subcaption}

\title{3D Gaze Estimation from 2D Pupil Positions on\\Monocular Head-Mounted Eye Trackers}

 \author{Mohsen Mansouryar \qquad Julian Steil \qquad Yusuke Sugano \qquad Andreas Bulling\\
 Perceptual User Interfaces Group\\
 Max Planck Institute for Informatics, Saarbr\"ucken, Germany\\\texttt{\{mohsen,jsteil,sugano,bulling\}@mpi-inf.mpg.de}}

\pdfauthor{Mohsen Mansouryar, Julian Steil, Yusuke Sugano and Andreas Bulling}

\TOGonlineid{183}


\setcopyright{rightsretained}

\conferenceinfo{The main part of this technical report was presented at ETRA 2016}{March 14 - 17, 2016, Charleston, SC, USA.}

\begin{document}

\conferenceinfo{The main part of this technical report was presented at ETRA 2016}{March 14 - 17, 2016, Charleston, SC, USA.}

 \teaser{
 \centering

 \includegraphics[width=0.3\textwidth]{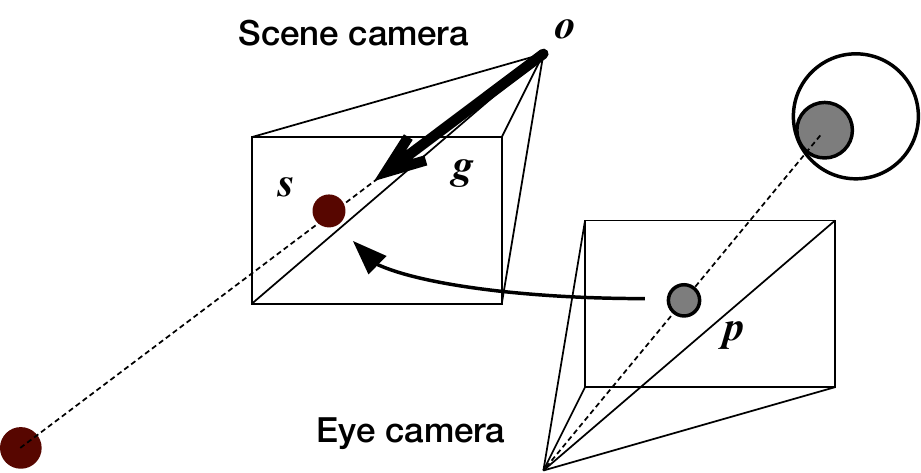}
 \includegraphics[width=0.3\textwidth]{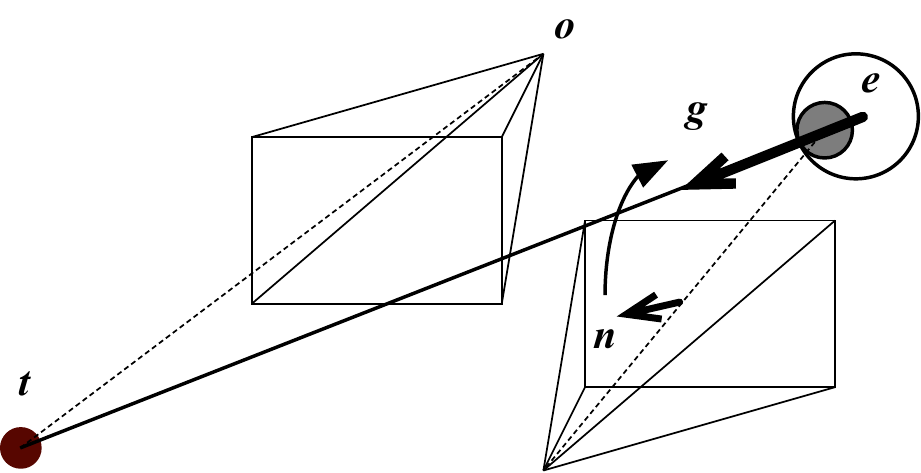}
 \includegraphics[width=0.3\textwidth]{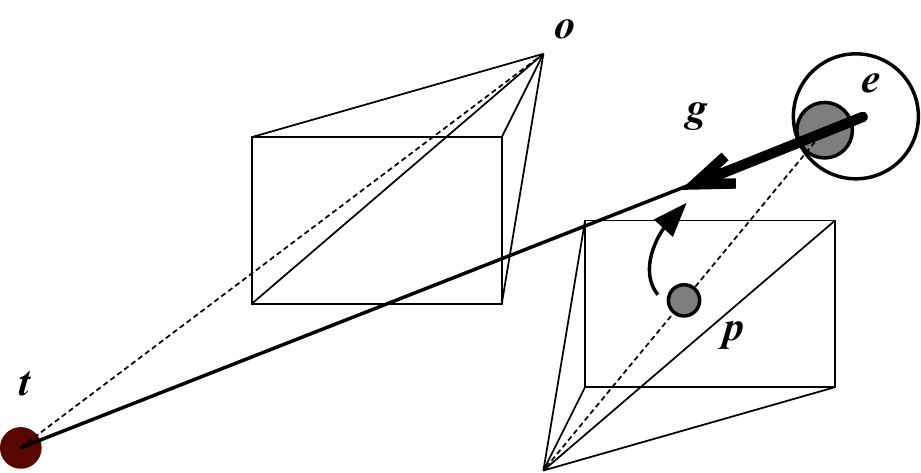}
 \\
 \mbox{\scriptsize\textbf{(a)} \textit{2D-to-2D mapping}} \hspace{3.0cm} \mbox{\scriptsize\textbf{(b)} \textit{3D-to-3D mapping}} \hspace{3.0cm} \mbox{\scriptsize\textbf{(c)} \textit{2D-to-3D mapping}}\
 \caption{Illustration of the (a) 2D-to-2D, (b) 3D-to-3D, and (c) 2D-to-3D mapping approaches. 3D gaze estimation in wearable settings is a task of inferring 3D gaze vectors in the scene camera coordinate system.}\label{fig:mappings}
 }

\maketitle

\begin{abstract}

3D gaze information is important for scene-centric attention analysis, but accurate estimation and analysis of 3D gaze in real-world environments remains challenging.
We present a novel 3D gaze estimation method for monocular head-mounted eye trackers.
In contrast to previous work, our method does not aim to infer 3D eyeball poses, but directly maps 2D pupil positions to 3D gaze directions in scene camera coordinate space.
We first provide a detailed discussion of the 3D gaze estimation task and summarize different methods, including our own.
We then evaluate the performance of different 3D gaze estimation approaches using both simulated and real data.
Through experimental validation, we demonstrate the effectiveness of our method in reducing parallax error, and we identify research challenges for the design of 3D calibration procedures.

\end{abstract}

\keywordlist

 \printcopyright

\section{Introduction}

Research on head-mounted eye tracking has traditionally focused on estimating gaze in screen coordinate space, e.g.\ of a public display.
Estimating gaze in scene or world coordinates enables gaze analysis on 3D objects and scenes and has the potential for new applications, such as real-world attention analysis~\cite{bulling16_computer}.
This approach requires two key components: 3D scene reconstruction and 3D gaze estimation.

In prior work, 3D gaze estimation was approximately addressed as a projection from estimated 2D gaze positions in the scene camera image to the corresponding 3D scene~\cite{munn20083d,takemura2014estimating,pfeiffer2014eyesee3d}.
However, without proper 3D gaze estimation, gaze mapping suffers from parallax error caused by the offset between the scene camera origin and eyeball position~\cite{mardanbegi2012parallax,duchowski2014comparing}.
To fully utilize the 3D scene information it is essential to estimate 3D gaze vectors in the scene coordinate system.

While 3D gaze estimation has been widely studied in remote gaze estimation, there have been very few studies in head-mounted eye tracking.
This is mainly because 3D gaze estimation typically requires model-based approaches with special hardware, such as multiple IR light sources and/or stereo cameras~\cite{beymer2003eye,nagamatsu2010user}.
Hence, it remains unclear whether 3D gaze estimation can be done properly only with a lightweight head-mounted eye tracker.
\'Swirski and Dodgson proposed a method to recover 3D eyeball poses from a monocular eye camera~\cite{Swirski2013}.
While it can be applied to lightweight mobile eye trackers, their method has been only evaluated with synthetic eye images, and its realistic performance including the eye-to-scene camera mapping accuracy has never been quantified.

We present a novel 3D gaze estimation method for monocular head-mounted eye trackers.
Contrary to existing approaches, we formulate 3D gaze estimation as a direct mapping task from 2D pupil positions in the eye camera image to 3D gaze directions in the scene camera.
Therefore, for the calibration we collect the 2D pupil positions as well as 3D target points, and finally minimize the distance between the 3D targets and the estimated gaze rays.

The contributions of this work are threefold.
First, we summarize and analyze different 3D gaze estimation approaches for a head-mounted setup.
We discuss potential error sources and technical difficulties in these approaches, and provide clear guidelines for designing lightweight 3D gaze estimation systems.
Second, following from this discussion, we propose a novel 3D gaze estimation method.
Our method directly maps 2D pupil positions in the eye camera to 3D gaze directions, and does not require 3D observation from the eye camera.
Third, we provide a detailed comparison of our method with state-of-the-art methods in terms of 3D gaze estimation accuracy.
The open-source simulation environment and the dataset are available at \url{http://mpii.de/3DGazeSim}.

\section{3D Gaze Estimation}

3D gaze estimation is the task of inferring 3D gaze vectors to the target objects in the environment.
Gaze vectors in scene camera coordinates can then be intersected with the reconstructed 3D scene.
There are three mapping approaches we discuss in this paper: 2D-to-2D, 3D-to-3D, and our novel 2D-to-3D mapping approach.
In this section, we briefly summarize three approaches. For more details, please refer to the appendix section.

\subsection*{2D-to-2D Mapping}

Standard 2D gaze estimation methods assume 2D pupil positions $\bm{p}$ in the eye camera images as input.
The task is to find the mapping function from $\bm{p}$ to 2D gaze positions $\bm{s}$ in the scene camera images (Figure 1 (a)).
Given a set of $N$ calibration data items ${(\bm{p}_i, \bm{s}_i)}_{i=1}^N$, the mapping function is typically formulated as a polynomial regression. 
2D pupil positions are first converted into their polynomial representations $\bm{q}(\bm{p})$, and the linear regression weight is obtained via linear regression methods.
Following Kassner et al.~\cite{Kassner14_ubicomp}, we did not include cubic terms and used an anisotropic representation as $\bm{q} = (1, u, v, uv, u^2, v^2, u^2 v^2)$ where $\bm{p} = (u, v)$. 

In order to obtain 3D gaze vectors, most of the prior work assumes that the 3D gaze vectors are originating from the origin of the scene camera coordinate system.
In this case, estimated 2D gaze positions $\bm{f}$ can be simply back-projected to 3D vectors $\bm{g}$ in the scene camera coordinate system.
This is equivalent to assuming that the eyeball center position $\bm{e}$ is exactly the same as the origin $\bm{o}$ of the scene camera coordinate system.
However, in practice there is always an offset between the scene camera origin and the eyeball position, and this offset causes the parallax error.

\subsection*{3D-to-3D Mapping}

If we can estimate a 3D pupil pose (unit normal vector of the pupil disc) from the eye camera as done in \'{S}wirski and Dodgson~\cite{Swirski2013}, we can instead take a direct 3D-to-3D mapping approach (Figure 1 (b)).
Instead of the 2D calibration targets $\bm{s}$, we assume 3D calibration targets $\bm{t}$ in this case.

With the calibration data ${(\bm{n}_i, \bm{t}_i)}_{i=1}^N$, the task is to find the rotation $\bm{R}$ and translation $\bm{T}$ between the scene and eye camera coordinate systems.
This can be done by minimizing distances between 3D gaze targets $\bm{t}_i$ and the 3D gaze rays which are rotated and translated to the scene camera coordinate system.
In the implementation, we further parameterize the rotation $\bm{R}$ by a 3D angle vector with the constraint that rotation angles are between $-\pi$ and $\pi$, and we initialize $\bm{R}$ assuming that the eye camera and the scene camera are facing opposite directions.

\subsection*{2D-to-3D Mapping}

Estimating 3D pupil pose is not a trivial task in real-world settings.
Another potential approach is to directly map 2D pupil positions $\bm{p}$ to 3D gaze directions $\bm{g}$ (Figure 1 (c)).

In this case, we need to map the polynomial feature $\bm{q}$ to unit gaze vectors $\bm{g}$ originating from an eyeball center $\bm{e}$.
$\bm{g}$ can be parameterized in a polar coordinate system,
and we assume a linear mapping from the polynomial feature $\bm{q}$ to the angle vector.
The regression weight is obtained by minimizing distances between 3D calibration targets $\bm{t}_i$ and the mapped 3D gaze rays as in the 3D-to-3D approach.
In the implementation, we used the same polynomial representation as the 2D-to-2D method to provide a fair comparison with the baseline.

\section{Data Collection}

In order to evaluate the potential and limitations of the introduced mapping approaches, we conducted two studies. First, we used data we obtained from a simulation environment, whereas the second study exploited real-world data collected from 14 participants.

\subsection*{Simulation Data}
\begin{figure}[t]
\centering
\includegraphics[width=1.0\linewidth]{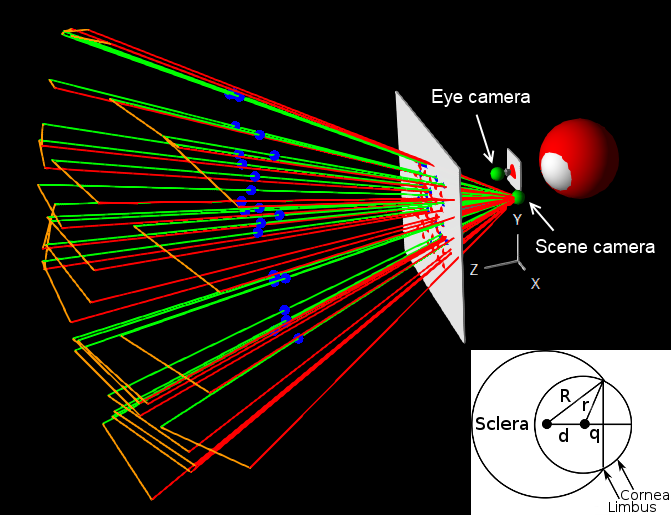}
\caption{3D eye model and simulation environment with 3D target points given as blue dots. The green and red rays correspond to ground truth and estimated gaze vectors, respectively.}
\label{fig:sim}
\end{figure}

We first analyzed the different mapping approaches in a simulation environment. 
Our simulation environment is based on a basic model of the human eye consisting of a pair of spheres~\cite{lefohn2003ocularist} and the scene and eye camera models. 
The eye model and a screenshot of the simulation environment are illustrated in \autoref{fig:sim}. We used human average anatomical parameters: $R = 11.5mm$, $r = 7.8mm$, $d = 4.7mm$, and $q = 5.8mm$.
The pupil is considered as the center of the circle which represents the intersection of the two spheres.
For both eye and scene cameras, we used the pinhole camera model. Intrinsic parameters were set to values similar to those of the actual eye tracking headset we used in the real-world environment.

One of the key questions about 3D gaze estimation is whether calibration at single depth is sufficient or not. Intuitively, obtaining calibration data at different depths from the scene camera can improve the 3D mapping performance.
We set calibration and test plane depths $d_c$ and $d_t$ to 1m, 1.25m, 1.5m, 1.75m, and 2m.
At each depth, points are selected from two grids, a 5 by 5 grid which gives us 25 calibration points (blue) and an inner 4 by 4 grid for 16 test points (red) displayed on the white plane of \autoref{fig:sim}. Both of the grids are symmetric with respect to the scene camera's principal axis. From the eye model used, we are able to estimate the corresponding gaze ray.

\subsection*{Real-World Data}

We also present evaluation of gaze estimation approaches using a real-world dataset to show the validity of 3D gaze estimation approaches.

\subsubsection*{Procedure}

\begin{figure}[t]

    \centering

    \begin{subfigure}[b]{0.515\columnwidth}
        \includegraphics[width=\columnwidth]{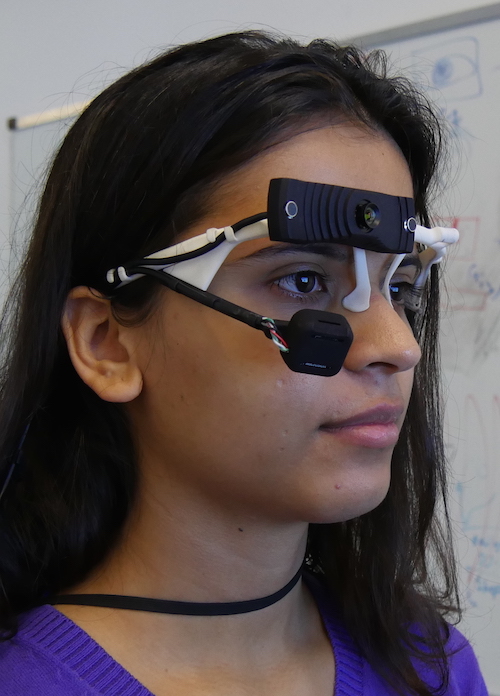}
        \caption{Video-based head-mounted eye tracker}
        \label{fig:tracker}
    \end{subfigure}
    \hspace{0.2cm}
    \begin{subfigure}[b]{0.4\columnwidth} 
        \includegraphics[width=\columnwidth]{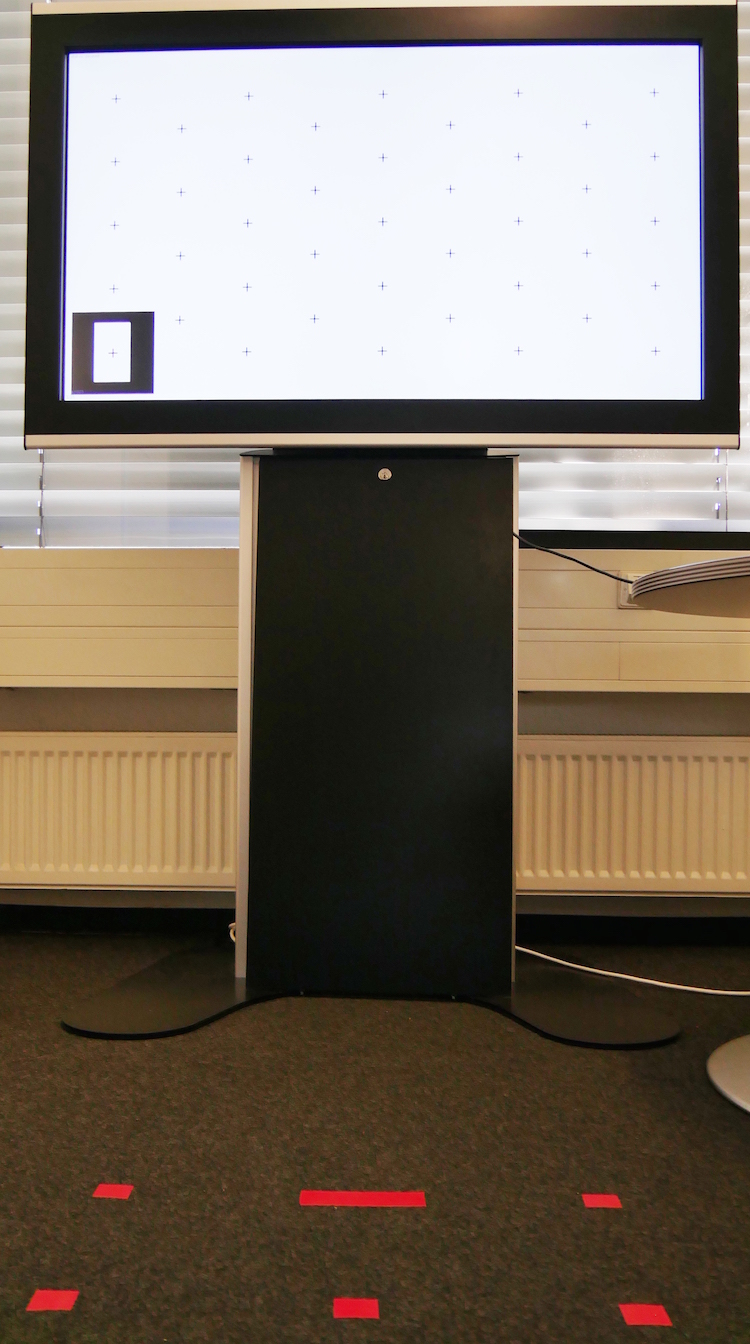}
        \caption{Display and distance markers}
        \label{fig:setup}
    \end{subfigure}
    \caption{The recording setup consisted of a Lenovo G580 laptop, a Phex Wall 55" display and a PUPIL head-mounted eye tracker.}
    \label{fig:recordingenvironment}
\end{figure}

The recording system consisted of a Lenovo G580 laptop and a Phex Wall 55" display (121.5cm $\times$ 68.7cm) with a resolution of 1920 $\times$ 1080.
Gaze data was collected using a PUPIL head-mounted eye tracker connected to the laptop via USB~\cite{Kassner14_ubicomp} (see Figure~\ref{fig:tracker}).
The eye tracker has two cameras: one eye camera with a resolution of $640 \times 360$ pixels recording a video of the right eye from close proximity, as well as an egocentric (scene) camera with a resolution of $1280 \times 720$ pixels.
Both cameras recorded videos at 30 Hz.
Pupil positions in the eye camera were detected using the PUPIL eye tracker's implementation.

We implemented remote recording software which conducts the calibration and test recordings shown on the display to the participants.
As shown in Figure~\ref{fig:setup}, the target markers were designed so that their 3D positions can be obtained using the ArUco library~\cite{Aruco2014}.
Intrinsic parameters of the scene and eye cameras were calibrated before recording, and used for computing 3D fixation target positions $\bm{t}$ and 3D pupil poses $\bm{n}$.

We recruited 14 participants aged between 22 and 29 years.
The majority of them had little or no previous experience with eye tracking.
Every participant had to perform two recordings, a calibration and a test recording of five different distances from the display.
Recording distances were marked by red stripes on the ground (see Figure~\ref{fig:setup}). They were aligned parallel to the display with an initial distance of 1 meter and the following recording distances with a spacing of 25cm (1.0, 1.25, 1.5, 1.75, 2.0). For every participant we recorded 10 videos.
 
As in the simulation environment, the participants were instructed to look at 25 fixation target points from the grid pattern in Figure~\ref{fig:setup}. 
After this step the participants had to perform the same procedure again while looking at 16 fixation targets placed on different positions than in the initial calibration to collect the test data for our evaluation part. 
This procedure was then repeated for the other four mentioned distances.
The only restriction we imposed was that the participants should not move their head during the recording.

\subsubsection*{Error Measurement}

Since the ground-truth eyeball position $\bm{e}$ is not available in the real-world study, we evaluate the estimation accuracy using an angular error observed from the scene camera.
For the case where 2D gaze positions are estimated (2D-to-2D mapping), 
we back-projected the estimated 2D gaze position $\bm{f}$ into the scene, and directly measured the angle $\theta$ between this line and the line from the origin of the scene camera $\bm{o}$ to the measured fixation target $\bm{t}$. 
For the cases where 3D gaze vectors are estimated, we first determined the estimated 3D fixation target position $\bm{t}'$ assuming the same depth as the ground-truth target $\bm{t}$.
Then the angle between the lines from the origin $\bm{o}$ was measured.

\section{Results}

\begin{figure}[t]
\centering
\includegraphics[width=1.0\linewidth]{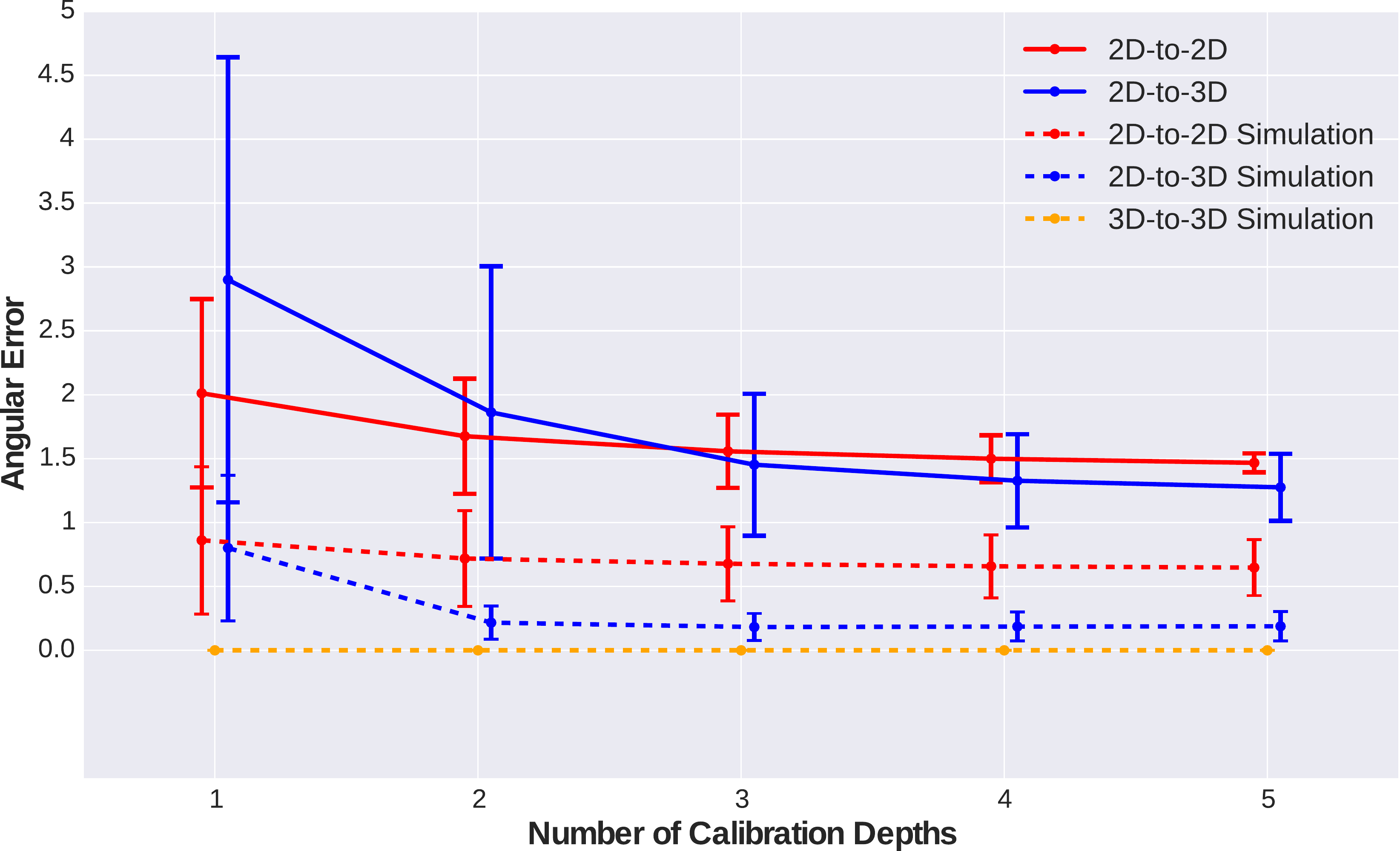}
\caption{Angular error performance over different numbers of calibration depths for 2D-to-2D, 2D-to-3D and 3D-to-3D mapping approaches. Each point corresponds to the mean over all angular error values for each number of calibration depths. The error bars provide the corresponding standard deviations.}
\label{fig:totalmean}
\end{figure}

We compared different mapping approaches in Figure~\ref{fig:totalmean} using an increasing number of calibration depths in both simulation and real-world environments. 
Each plot corresponds to mean estimation errors of all test planes and all combinations of calibration planes.
Angular error is evaluated from the ground-truth eyeball position.
It can be seen that in all cases the estimation performance can be improved by taking more calibration planes. Even the 2D-to-2D mapping approach performs slightly better with multiple calibration depths overall in both environments. 
The 2D-to-3D mapping approach performed better than the 2D-to-2D mapping in all cases in the simulation environment.
For the 3D-to-3D mapping approach a parallax error near to zero can be achieved.

Similarly to the simulation case, we first compare the 2D-to-3D mapping with the 2D-to-2D mapping in terms of the influence of different calibration depths displayed as stable lines in Figure~\ref{fig:totalmean}. 
Since it turned out that the 3D-to-3D mapping on real-world data has more angular error (over 10$^\circ$) than the 2D-to-3D mapping, we omit the results in the following analysis.

Contrary to the simulation result, with a lower number of calibration depths the 2D-to-2D approach performs better than the 2D-to-3D approach for real-world data. However, with an increasing number of calibration depths, the 2D-to-3D approach outperforms 2D-to-2D comparing the angular error in visual degrees. For five calibration depths we can achieve for the 2D-to-3D case an overall mean of less than 1.3 visual degrees over all test depths and all participants. 
A more detailed analysis and discussion with corresponding performance plots are available in the appendix.

\section{Discussion}

We discussed three different approaches for 3D gaze estimation using head-mounted eye trackers.
Although it was shown that the 3D-to-3D mapping is not a trivial task, the 2D-to-3D mapping approach was shown to perform better than the standard 2D-to-2D mapping approach using simulation data.
One of the key observations from the simulation study is that the 2D-to-3D mapping approach requires at least two calibration depths.
Given more than two calibration depths, the 2D-to-3D mapping can significantly reduce the parallax error.

On the real data, we could observe a decreasing error for the 2D-to-3D mapping with an increasing number of calibration depths, and could outperform the 2D-to-2D mapping.
However, the performance of the 2D-to-3D mapping became worse than in the simulation environment.
Reasons for the different performance of the mapping approaches in the simulation and real-world environment are manifold and reveal their limitations. 
Our simulation environment considers an ideal setting and does not include noise that occurs in the real world. 
This noise is mainly produced by potential errors in the pupil and marker detection, as well as head movements of the participants. 

In future work it will be important to investigate how the 3D-to-3D mapping approach can work in practice.
The fundamental difference from the 2D-to-3D mapping is that the mapping function has to explicitly handle the rotation between eye and scene camera coordinate systems.
In addition to the fundamental estimation inaccuracy of the 3D pupil pose estimation technique without modeling real-world factors such as corneal refraction, we did not consider the difference between optical and visual axes.
A more appropriate mapping function could be a potential solution for the 3D-to-3D mapping, and another option could be to use more general regression techniques considering the 2D-to-3D results.

Throughout the experimental validation, this research also illustrated the fundamental difficulty of the 3D gaze estimation task.
It has been shown that the design of the calibration procedure is also quite important, and it is essential to address the issue from the standpoint of both calibration design and mapping formulation.
Since the importance of different calibration depths has been shown, the design of automatic calibration procedure, e.g., how to obtain calibration data at different depths using only digital displays, is another important HCI research issue.

Finally, it is also important to combine the 3D gaze estimation approach with 3D scene reconstruction methods and evaluate the overall performance of 3D gaze mapping.
In this sense, it is also necessary to evaluate performance with respect to scene reconstruction error.

\section{Conclusion}

In this work, we provided an extensive discussion on different approaches for 3D gaze estimation using head-mounted eye trackers.
In addition to the standard 2D-to-2D mapping approach, we discussed two potential 3D mapping approaches using either 3D or 2D observation from the eye camera.
We conducted a detailed analysis of 3D gaze estimation approaches using both simulation and real data.

Experimental results showed the advantage of the proposed 2D-to-3D estimation methods, but its complexity and technical challenges were also revealed.
Together with the dataset and simulation environment, this study would provide a solid basis for future research on 3D gaze estimation with lightweight head-mounted devices.

\section*{Acknowledgements}

We would like to thank all participants for their help with the data collection. This work was funded, in part, by the Cluster of Excellence on Multimodal Computing and Interaction (MMCI) at Saarland University, the Alexander von Humboldt Foundation, and a JST CREST research grant.

\bibliographystyle{acmsiggraph}
\bibliography{template}

\clearpage

\section*{Appendix}

\subsection*{3D Gaze Estimation Approaches}

We first introduce detailed formulations of three approaches that are briefly presented above.

\subsubsection*{2D-to-2D Mapping Approach}

As briefly described above, standard 2D gaze estimation methods assume 2D pupil positions $\bm{p}$ in the eye camera images as input, and the task is to find the polynomial mapping function from $\bm{p}$ to 2D gaze positions $\bm{s}$ in the scene camera images.
2D pupil positions are first converted into their polynomial representations $\bm{q}(\bm{p})$, and a coefficient vector $\bm{w}$ which minimizes a cost function
\begin{eqnarray}
 E_{\textmd{2Dto2D}}(\bm{w}) = \sum_{i=1}^N | \bm{s}_i - \bm{q}_i\bm{w}|^2
 \end{eqnarray}
is obtained via linear regression methods.
Then any pupil positions $\bm{p}$ can be mapped to 2D gaze positions as
$\bm{f} = \bm{q}\bm{w}$.

\subsubsection*{3D-to-3D Mapping Approach}

In this case, the input to the mapping function is 3D pupil pose unit vectors $\bm{n}$.
Given the calibration data ${(\bm{n}_i, \bm{t}_i)}_{i=1}^N$ with 3D calibration targets $\bm{t}$, the task is to find the rotation $\bm{R}$ and translation $\bm{T}$ between the scene and eye camera coordinate systems.

If we denote the origin of the pupil pose vectors as $\bm{e}_{\textmd cam}$, 3D gaze rays after the rotation and translation are defined as a line $\bm{e}_{\textmd cam} + \bm{T} + \lambda \bm{R} \bm{n}$, where $\lambda$ parameterize the gaze line\footnote{Please note that $\lambda$ is the parameter required to determine the 3D gaze point by intersecting the gaze ray to the scene, and does not have to be obtained during calibration stage.}.
Given the calibration data, $\bm{R}$ and $\bm{T}$ are obtained by minimizing distances $d_i$ between 3D gaze targets $\bm{t}_i$ and the 3D gaze rays.
In a vector form, the squared distance $d_i^2$ can be written as
\begin{eqnarray}
d_i^2 &=& \frac{| \bm{R} \bm{n}_i  \times (\bm{t}_i - (\bm{e}_{\textmd cam} + \bm{T})) |^2}{| \bm{R} \bm{n}_i |^2} \nonumber \\
&=& | \bm{R} \bm{n}_i  \times (\bm{t}_i - (\bm{e}_{\textmd cam} + \bm{T})) |^2.
\end{eqnarray}
Since $\bm{e}_{\textmd cam} + \bm{T}$ denotes the eyeball center position $\bm{e}$ in the scene camera coordinate system, the cost function can be defined as
\begin{eqnarray}
 E_{\textmd{3Dto3D}}(\bm{R}, \bm{e}) = \sum_{i=1}^N |\bm{R} \bm{n}_i  \times (\bm{t}_i - \bm{e}) |^2.\label{eq:3Dto3Dcost}
\end{eqnarray}
Minimization of Eq.~(\ref{eq:3Dto3Dcost}) can be done using nonlinear optimization methods such as the Levenberg-Marquardt algorithm. 
At the initialization step of the nonlinear optimization, we assume $\bm{e}_0=(0, 0, 0)$ and $\bm{R}_0=(0, \pi, 0)$ considering the opposite direction of the scene and eye cameras in the world coordinate system. 

\subsubsection*{2D-to-3D Mapping Approach}

Another potential approach is to directly map 2D pupil positions $\bm{p}$ to 3D gaze directions $\bm{g}$.
In this case, we map the polynomial feature $\bm{q}$ to unit gaze vectors $\bm{g}$ originating from the eyeball center $\bm{e}$ in the scene camera coordinate system.
$\bm{g}$ can be parameterized in a polar coordinate system as
 \begin{eqnarray}
 \bm{g} = \begin{pmatrix}
 \sin \theta \\
 \cos \theta \sin\phi\\
 \cos \theta \cos\phi
 \end{pmatrix},
 \end{eqnarray}
 and we assume a linear mapping from the polynomial feature $\bm{q}$ to the angle vector as
 \begin{eqnarray}
 \bm{\alpha} = (\theta, \phi) = \bm{q}\bm{w}.
 \end{eqnarray}

Given the 3D calibration data ${(\bm{p}_i, \bm{t}_i)}_{i=1}^N$, $\bm{w}$ can be obtained by minimizing distances $d_i$ between 3D gaze targets $\bm{t}_i$ and the gaze rays.
Therefore, similarly to the 3D-to-3D mapping case, the target cost function to be minimized is
\begin{eqnarray}
E_{\textmd{2Dto3D}}(\bm{w}, \bm{e}) = \sum_{i=1}^N |\bm{g}(\bm{q}_i\bm{w}) \times (\bm{t}_i - \bm{e})|^2.\label{eq:2Dto3Dcost}
\end{eqnarray}

In order to initialize the parameters for nonlinear optimization, we first set $\bm{e}_0=(0, 0, 0)$.
Then using the polar coordinates of gaze targets $\bm{t}_i=(\theta_i, \phi_i)$, the initial $\bm{w_0}$ can be obtained by solving the linear regression problem
\begin{eqnarray}
E(\bm{w}) = \sum_{i=1}^N | \bm{(\theta_i, \phi_i)} - \bm{q}_i\bm{w}|^2.
\end{eqnarray}

\subsection*{Extended Analysis}

In this section, we provide extended analysis on the different performance taking single and multiple calibration depth combinations into account.

\subsubsection*{Simulation Study}

\begin{figure*}[t]
    \centering
    \begin{subfigure}[b]{0.49\textwidth}
        \includegraphics[width=\textwidth]{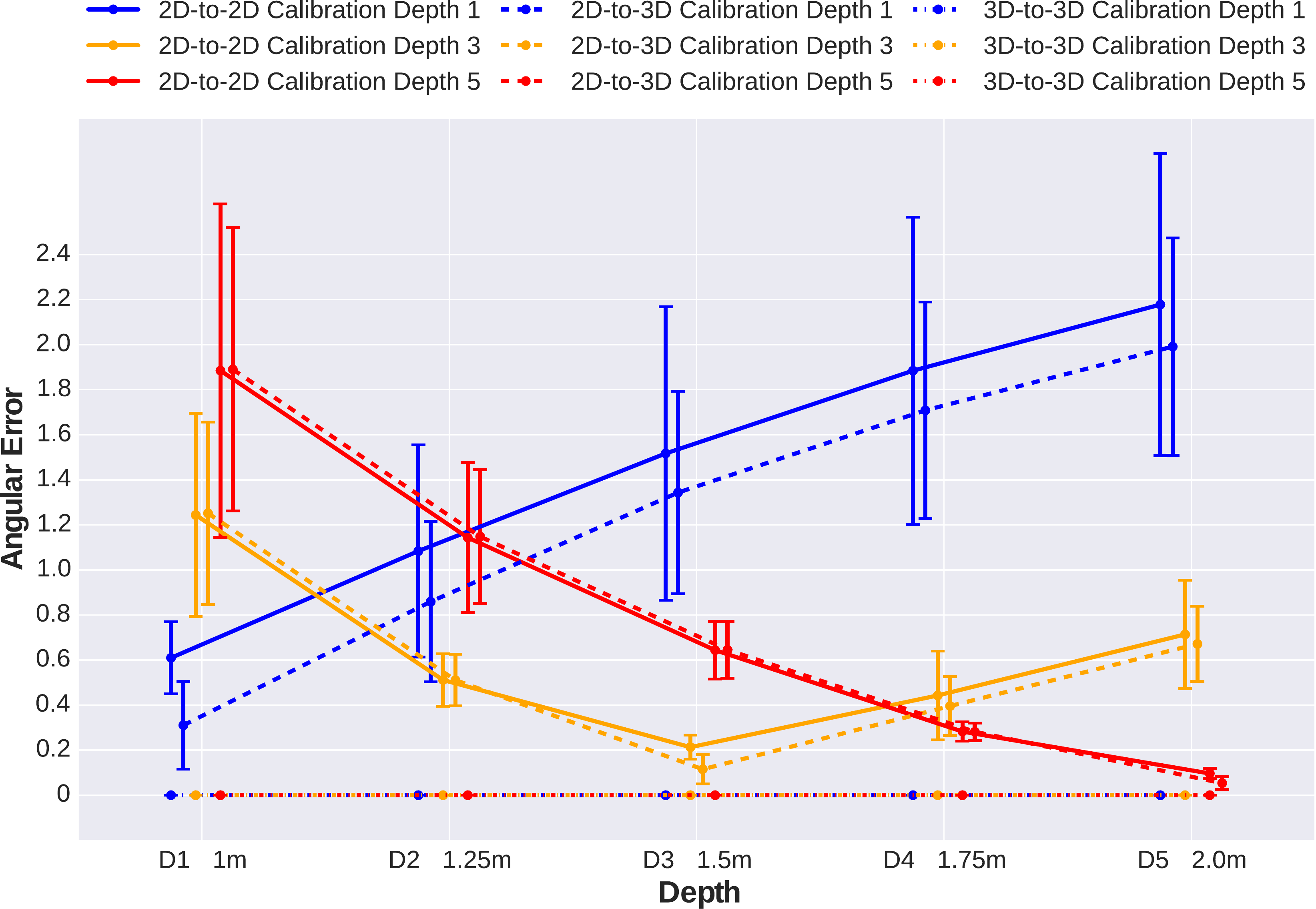}
        \caption{One calibration depth setting for 2D-to-2D, 2D-to-3D and 3D-to-3D mappings}
        \label{fig:parallax_comparison}
    \end{subfigure}
    \begin{subfigure}[b]{0.49\textwidth}
        \includegraphics[width=\textwidth]{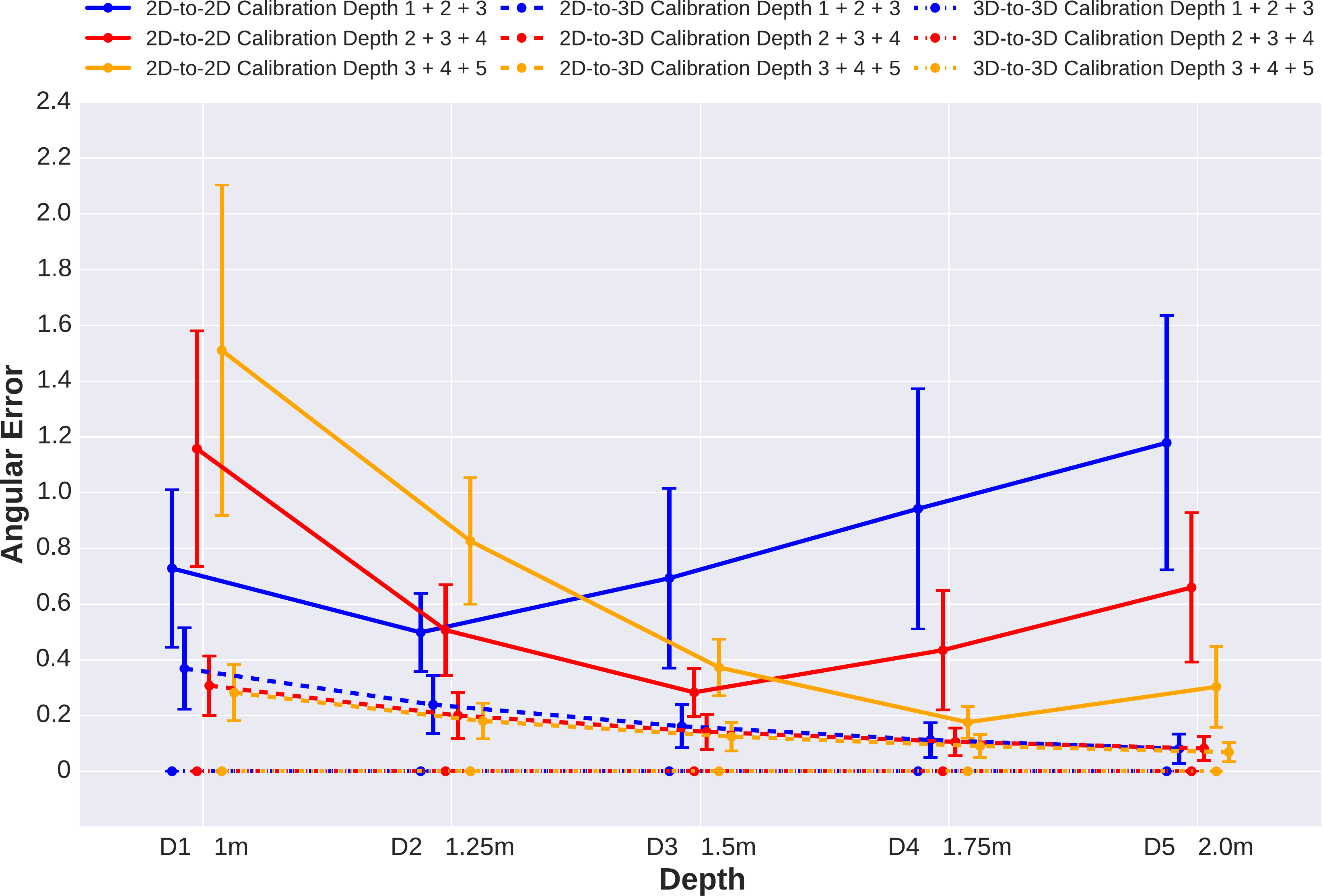}
        \caption{Three calibration depth setting for 2D-to-2D, 2D-to-3D and 3D-to-3D mappings}
        \label{fig:parallax_comparison_3depths}
    \end{subfigure}

    \caption{Comparison of parallax error. Vertical axis shows the mean angular error values over all test depths (D1-D5). Dashed lines correspond to the 2D-to-3D mapping, Dotted lines correspond to 3D-to-3D and solid lines correspond to the 2D-to-2D mapping. Each color represents one of the different calibration depth settings.}
    \label{fig:parallax_comparison_sim}
\end{figure*}

\begin{figure*}[t]
    \centering
    \begin{subfigure}[b]{0.49\textwidth}
        \includegraphics[width=\textwidth]{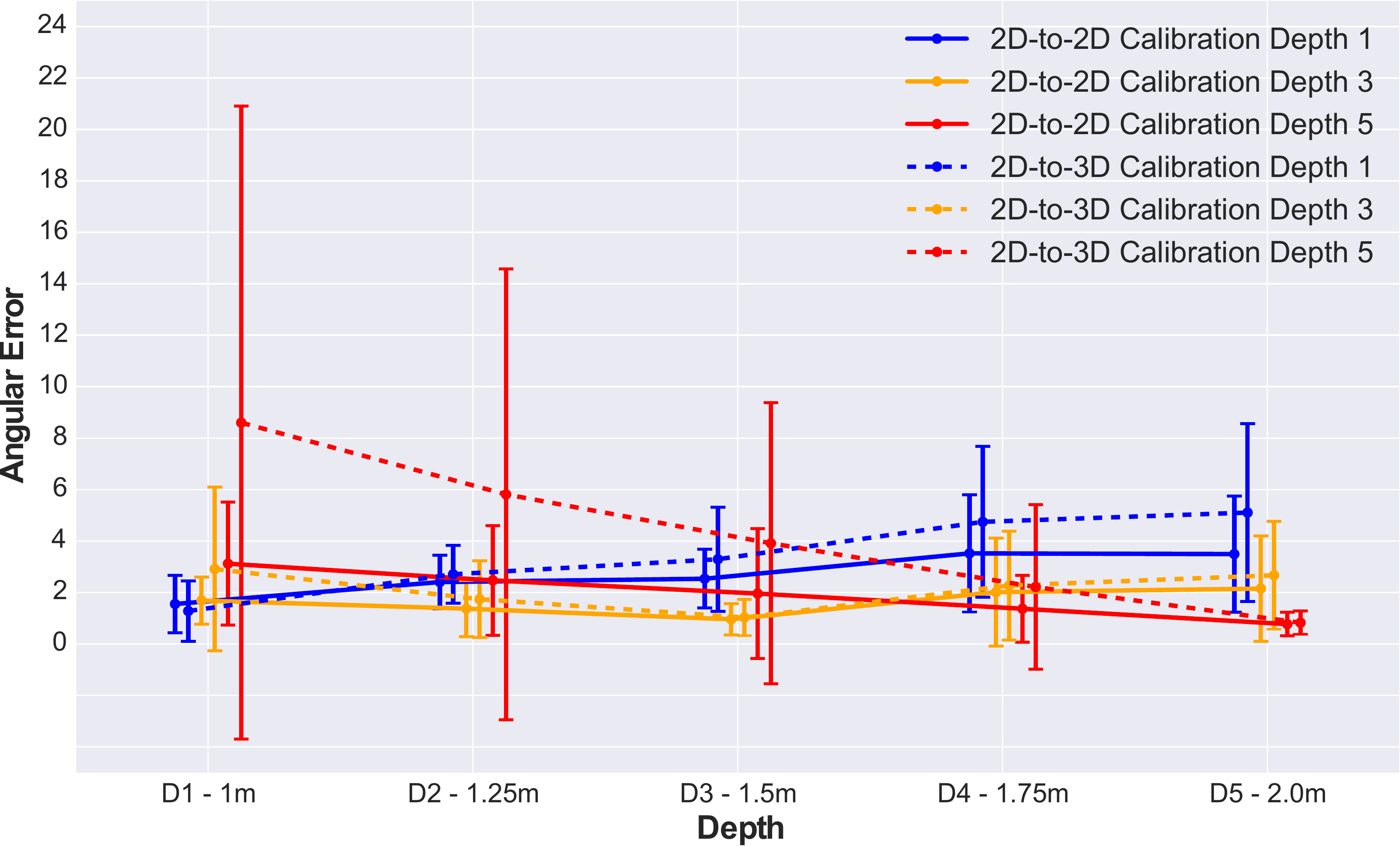}
        \caption{One calibration depth setting for 2D-to-2D and 2D-to-3D mappings}
        \label{fig:1D}
    \end{subfigure}
    \begin{subfigure}[b]{0.49\textwidth}
        \includegraphics[width=\textwidth]{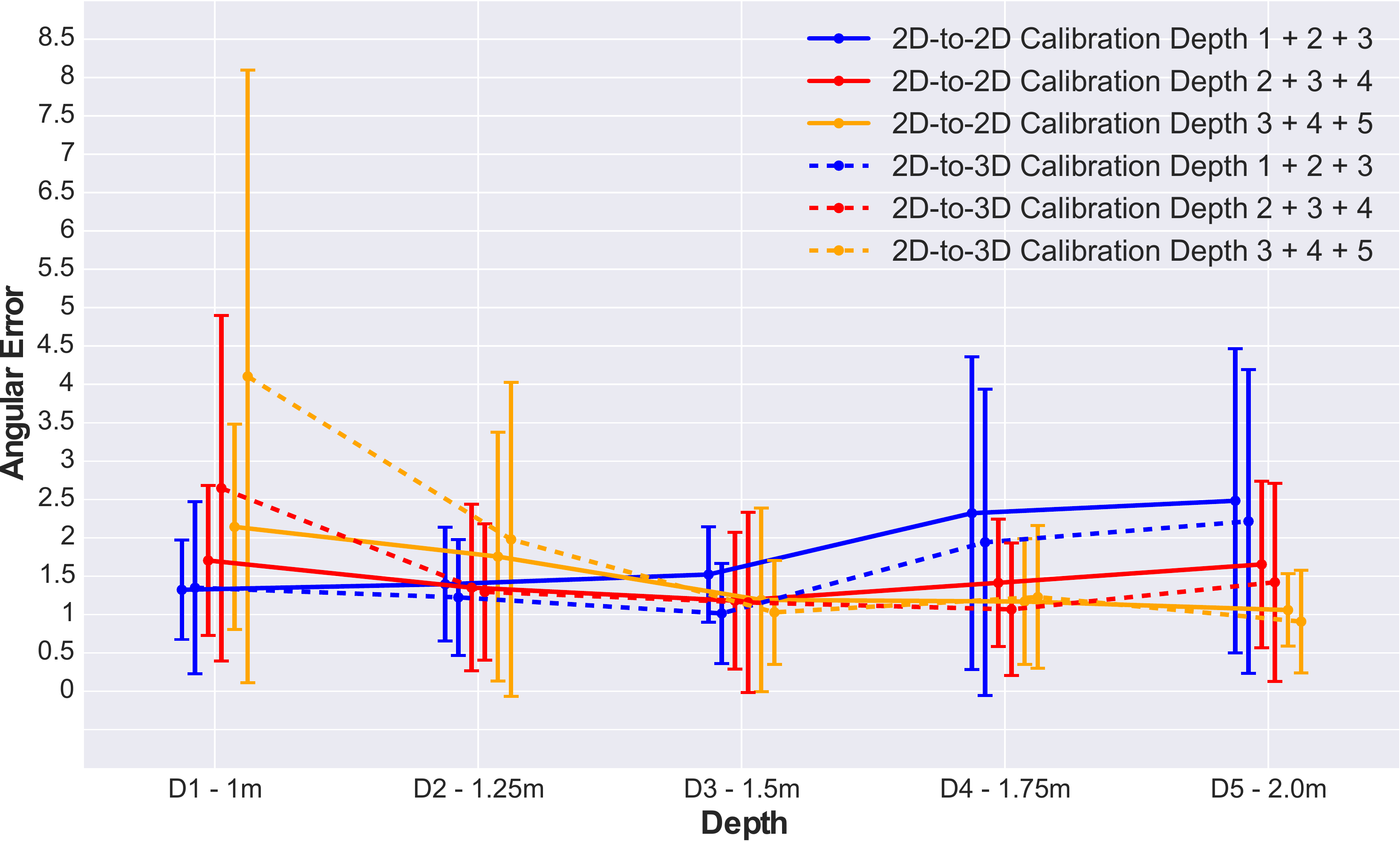}
        \caption{Three calibration depth setting for 2D-to-2D and 2D-to-3D mappings}
        \label{fig:3D}
    \end{subfigure}

    \caption{Comparison of one and three calibration depths considering the mean angular error values over all participants and for every test depth (D1-D5). The error bars show the corresponding standard deviation for every test depth.}
    \label{fig:calibration_depth_comparison}
\end{figure*}

\autoref{fig:parallax_comparison_sim} shows the error for all three mapping approaches on the simulation data by fixing the calibration depth in a similar manner as in Mardanbegi and Hansen's work~\cite{mardanbegi2012parallax}. 
\autoref{fig:parallax_comparison} and \autoref{fig:parallax_comparison_3depths} are corresponding to  performances using one and three calibration depth, respectively.
Each plot shows the mean angular error distribution over test depths, and each color corresponds to a certain calibration depth. The error bars describe the corresponding standard deviations. 
Dashed lines correspond to the 2D-to-3D mapping, dotted lines correspond to the 3D-to-3D mapping, and solid lines correspond to the 2D-to-2D mapping.

With one calibration depth (\autoref{fig:parallax_comparison}), the performance of the 2D-to-3D mapping is always better than the 2D-to-2D case.
However, we can observe that the parallax error is still present in the 2D-to-3D case, which indicates the fundamental limitations of the approximated mapping approach.
With three calibration depth (\autoref{fig:parallax_comparison_3depths}), the 2D-to-3D mapping approach performs significantly better than in \autoref{fig:parallax_comparison} and the parallax error reaches a near zero level. However, there is a tendency for the error to become larger as the test depth becomes closer to the camera, which indicates the limitations of the proposed mapping function.
The performance of the 2D-to-2D mapping is also improved, but we can see that the increased number of calibration depths cannot be a fundamental solution to the parallax error issue.
For the 3D-to-3D mapping, the angular error is close to zero even for only one calibration depth. Taking more calibration depths into account does not lead to a further improvement.

\subsubsection*{Real-World Study}

Similarly, we show a detailed comparison of the 2D-to-2D and 2D-to-3D mapping approaches using the real-world data.
\autoref{fig:1D} displays the mean angular error for both approaches taking only one calibration depth over all 14 participants in the same manner as in \autoref{fig:parallax_comparison}.
For both mapping approaches, each calibration depth setting performed best for the corresponding test depth, and the error increased with an increased test distance from the calibration depth. 
However, for the 2D-to-2D approach the angular error values over all distances are smaller than for the 2D-to-3D case, except for the case where the calibration depth and test depth are the same. 

This behavior changes for an increasing number of calibration depths, as can be seen in Figure~\ref{fig:3D}, where we used three different calibration depths as in \autoref{fig:parallax_comparison_3depths}.
The 2D-to-3D mapping approach performs better than the 2D-to-2D mapping for nearly all combinations, except for the test depth D1, exploiting the additional 3D information collected during calibration to improve the gaze direction estimation.

\begin{figure}[h]
    \centering
    \includegraphics[width=1.0\columnwidth]{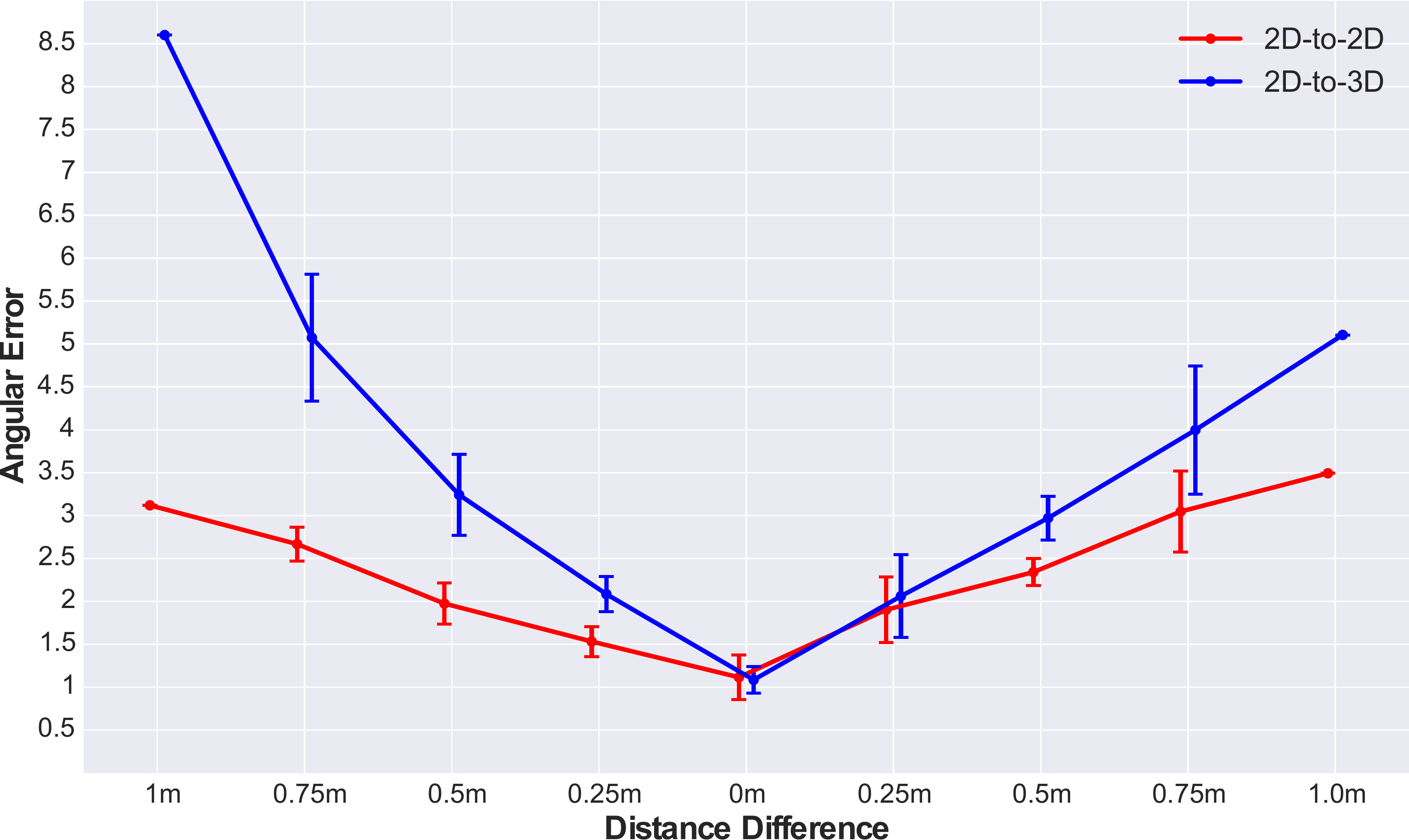}
    \caption{The effect of distance from the calibration depth for 2D-to-2D and 2D-to-3D, taking one calibration depth into account. Every point describes the mean angular error with respect to the offset between the calibration and test depth. The error bars provide the corresponding standard deviation.}
    \label{fig:calibration_distance}
\end{figure}

Figure~\ref{fig:calibration_distance} shows the mean angular errors with respect to the offset between the calibration and test depths for the one calibration depth setting.
The negative distance values on the horizontal axis indicate cases where the test depth is closer than the calibration depth, and vice versa for the positive distance values. 
As can be seen, the 2D-to-3D mapping approach tends to produce higher error if the test depth distance from the calibration depth increases.

\end{document}